\documentclass{aa}  
\usepackage{graphicx}
\usepackage{natbib}

\usepackage{txfonts}
\begin{document}
  \title{New X-ray observations of IQ Aurigae and $\alpha^{2}$\,Canum Venaticorum}
   \subtitle{Probing the magnetically channelled wind shock model in A0p stars}

   \author{J. Robrade
          \and
          J.H.M.M. Schmitt
          }


        \institute{Hamburger Sternwarte, Universit\"at Hamburg, Gojenbergsweg 112, 21029 Hamburg, Germany\\
       \email{jrobrade@hs.uni-hamburg.de}
             }

   \date{Received 07 March 2011; Accepted}

 
  \abstract
{}
{We re-examine the scenario of X-ray emission from magnetically confined/channelled wind shocks (MCWS) for Ap/Bp stars, 
a model originally developed to explain the $ROSAT$ detection of the A0p star IQ~Aur.}
{We present new X-ray observations of the A0p stars $\alpha^{2}$\,CVn ($Chandra$) and IQ~Aur ($XMM-Newton$)
and discuss our findings in the context of X-ray generating mechanisms of magnetic, chemically peculiar, intermediate mass stars.}
{The X-ray luminosities of IQ~Aur with $\log L_{\rm X} = 29.6$~erg\,s$^{-1}$ and $\alpha^{2}$\,CVn with $\log L_{\rm X} \lesssim 26.0$~erg\,s$^{-1}$  
differ by at least three orders of magnitude, despite both being A0p stars. 
By studying a sample of comparison stars, we find that X-ray emission is preferably generated by more massive objects such as IQ~Aur.
Besides a strong, cool plasma component, significant amounts of hot ($>$\,10~MK) plasma are present during the quasi-quiescent phase of IQ~Aur; further,
diagnostics of the UV sensitive $f/i$ line ratio in He-like \ion{O}{vii} triplet point to X-ray emitting regions well above the stellar surface of IQ~Aur.
In addition we detect a large flare from IQ~Aur with temperatures up to $\sim 100$~MK and a peak X-ray luminosity of $\log L_{\rm X} \approx 31.5$~erg\,s$^{-1}$.
The flare, showing a fast rise and e-folding decay time of less than half an hour, originates from a rather compact structure
and is accompanied by a significant metallicity increase. 
The X-ray properties of IQ~Aur cannot be described by wind shocks only and require the presence of magnetic reconnection. This is most evident
in the, to our knowledge, first X-ray flare reported from an A0p star.}
{Our study indicates that the occurrence the of X-ray emission in A0p stars generated by magnetically channelled wind shocks depends on stellar properties
such as luminosity which promote a high mass loss rate, whereas magnetic field configuration and transient phenomena refine their appearance.
While we cannot rule out unknown close companions, the X-ray emission from IQ~Aur can be
described consistently in the MCWS scenario, in which the very strong magnetic confinement of the stellar wind
has led to the build up of a rigidly rotating disk around the star, where magnetic reconnection and centrifugal breakout events occur.}
   \keywords{Stars: activity -- Stars-- chemically peculiar -- Stars: individual $\alpha^{2}$\,CVn, IQ Aur -- X-rays: stars
               }

   \maketitle
%

\section{Introduction}

X-ray emission is found virtually everywhere among the population of the main-sequence stars.
The X-ray generating mechanisms differ however for 'hot' O and early-B stars, and 'cool', late-A to M-type stars.
In hot stars the correlation between X-ray and bolometric luminosity  at $\log L_{\rm X}/L_{\rm bol} \approx -7$ \citep{berg97} 
is explained by X-ray emission arising from
shocks in instabilities in their radiatively driven winds. In cool stars,
X-ray emission originates in magnetic structures generated by dynamo processes, powered by an interplay of convective motions 
in the outer layers and stellar rotation.
The observed activity levels range from $\log L_{\rm X}/L_{\rm bol} \approx -7~...-3$ \citep{schmitt97}, whereas
an activity-rotation relation is present until the dynamo saturates \citep[for an overview on stellar coronae see][]{gue04}.
In intermediate mass stars, around spectral type late-A,
the outer convection zone becomes increasingly thinner and the dynamo
efficiency declines strongly. A prominent example is the A7 star Altair, 
one of the hottest magnetically active stars detected in X-rays \citep{rob09a}.
Thus one expects a virtually X-ray dark population at spectral types mid-B to mid-A, 
where stars neither possess an outer convection zone nor drive strong stellar winds.
The overall X-ray detection rate among those stars is with $\sim 10$\,\% indeed quite low, but
surprisingly X-ray emission was detected from several of these stars \citep{schroeder07}.
Consequently, low-mass stars hidden in the vicinity of their optically bright primaries have been often proposed
as the true origin of the X-ray emission.
Unresolved companions are a likely explanation for the X-ray emission from 'normal' main-sequence stars at these spectral types
and in some cases they even can be shown to be responsible for the X-ray emission, e.g. in the eclipsing binary systems like $\alpha$\,CrB \citep{schmitt93}.
However, especially
peculiar Ap/Bp stars and young Herbig AeBe stars are prime candidates for being intrinsic X-ray emitters \citep[e.g.][]{bab97,ste09}.
In the here discussed Ap/Bp stars an interplay between stellar winds and large scale magnetic fields is thought to play the major role in the X-ray generation.

Ap/Bp stars are magnetic, chemically peculiar (CP) stars that have completed an often significant fraction of their main sequence lifetime, 
i.e. they are not especially young.
Only a small fraction of about 5\,\% of the mid-B to mid-A type stars are magnetic;
these stars typically exhibit various chemical peculiarities and
a rather slow rotation compared to non-magnetic stars of similar spectral type.
The origin of their magnetic field is debated, but often thought to be of fossil origin,
consistent with the finding that
only a small fraction is magnetic and that field
strength and rotation are independent.
The magnetic field geometry in Ap/Bp stars is dominated by rather simple, large scale structures like
a dipole and hence it is fundamentally different from the complex field geometry
of magnetically active late-type stars \citep[see e.g. the overview by][]{lan92}.

The prototypical case for a presumably single, magnetic Ap star with detected X-ray emission is IQ~Aur.
In a 10~ks pointed observation with {\it ROSAT} PSPC, IQ~Aur was shown to be an 
X-ray source with a high X-ray luminosity of $L_{\rm X}= 4 \times 10^{29}$~erg\,s$^{-1}$ and
a quite soft spectrum with $T_{\rm X}=0.3$~keV \citep[][BM97]{bab97}. Thus a hypothetical late-type companion was
a rather unlikely explanation. In order
to explain the X-ray emission from IQ~Aur and, more generally of Ap/Bp stars, 
BM97 introduced the 'magnetically confined wind-shock model' (MCWS).
In the MCWS model the radiatively driven wind components from both hemispheres of the star are magnetically
confined and forced to collide in the vicinity of its equatorial plane. As a consequence, strong shocks form and
give rise to very efficient X-ray production from kinetic wind energy.
Although some fine-tuning is required to obtain a rather high mass loss rate,
the X-ray emission of IQ~Aur can actually be accounted for with reasonable assumptions on the wind parameters;
specifically BM97 assumed a stellar wind with $V_{\infty}=800$~km\,s$^{-1}$ and a mass loss rate of $\dot M_{\odot}=10^{-10} -10^{-11} M_{\odot}$\,yr$^{-1}$.

Notable advancements of the static BM97 model were achieved by magnetohydrodynamic simulations of 
magnetically channelled line-driven stellar winds \citep{dou02}. In a further study, these MHD simulations
investigate an extended parameter space and include stellar rotation in the modeling \citep{dou08}.
Dynamic variants of the MCWS model were successfully applied to Herbig AeBe and O-type stars and
may actually extend to a whole class of magnetic early-type stars. In this respect
it can be considered as a 'standard' model that is invoked to explain a variety of phenomena like
X-ray overluminosity, hard spectral components, flares or rotational modulation.
However, recent X-ray studies of Ap/Bp stars often remained inconclusive.
\cite{ste06a} studied X-ray detected intermediate mass main-sequence stars (B5-A0) with {\it Chandra}
including three CP stars, but
they do not find strong support for the MCWS model, especially due to the non-detection of the magnetic Ap star HD~133880.
\cite{cze07} studied mid-B to mid-A magnetic stars; they detected X-ray emission from the A0p star GL~Lac ( Babcock's star), though with
a much higher X-ray temperature and luminosity than IQ~Aur, but also report the non-detection of the A0p star HD~184905.
They find that the presence of a magnetic field
is not a sufficient criterion for X-ray emission; further X-ray luminosity and field strength are not correlated.

These findings make it desirable to investigate the MCWS scenario in its 'original' environment and
we present contemporary X-ray observations of two prototypical A0p stars, $\alpha^{2}$\,CVn and IQ~Aur. Our paper is structured as follows:
we summarize the targets properties in Sect.\,\ref{tar},
in Sect.\,\ref{ana} we describe observations and data analysis, present our findings
in Sect.\,\ref{res} and discuss them in a broader contexts in Sect.\ref{dis}.

\section{The targets $\alpha^{2}$\,CVn and IQ Aur}
\label{tar}

The A0p star $\alpha^{2}$\,CVn (\object{HD 112413}, HR~4915) is the primary component (V\,=\,2.9)
in a wide binary (angular separation 21.3\,\arcsec) with the F0V star HR~4914 as secondary.
$\alpha^{2}$\,CVn is the prototype of a class of spectrum and brightness variable stars, caused by the interplay of stellar rotation and starspots, i.e.
metal inhomogeneities in the stellar atmosphere.
These stars have spectral classifications from late-B to mid-A type and belong to the class of chemically peculiar, magnetic stars.
\cite{koch10} study spectropolarimetric data and find that the overall magnetic structure of $\alpha^{2}$\,CVn can be modelled with an 4.6~kG dipolar field, 
an inclination of $i \approx 120^{\circ}$ and an obliquity, i.e. the angle between the rotational
and the magnetic axes, of $\beta \approx 80^{\circ}$. Nevertheless, also higher complexity in the magnetic topology is present.
X-ray emission from this system was detected with {\it Einstein} (0.15\,--\,4.0~keV) at a level of $\log L_{\rm X} = 28.8$~erg\,s$^{-1}$ \citep{schmitt85}
and {\it ROSAT} (0.1\,--\,2.4~keV) at a level of $\log L_{\rm X} = 28.6$~erg\,s$^{-1}$ \citep{schroeder07}, 
but it was not resolved in both observations and the contribution from the individual components remained unknown.

The field A0p star IQ~Aur (\object{HD 34452}, HR~1732)
belongs to the class of $\alpha^2$\,CVn spectrum variable stars and is thought to be a single star with V=5.4 and small $E_{\rm B-V} \simeq 0.01$. 
BM97, who report the {\it ROSAT} X-ray detection at $\log L_{\rm X} = 29.6$~erg\,s$^{-1}$, assume a 4~kG dipolar field for IQ~Aur 
and determine an inclination of $i \approx 32^{\circ}$ and an obliquity
of $\beta \approx 50^{\circ}$,  whereas \cite{hub07a} find $i=87^{\circ}$ and of $\beta =7^{\circ}$.
\cite{boh93} state: 'a large range of values for $i$ is possible: if $i$ is large $\beta$ is fairly small and if $i$ is as small as allowed by our assumptions, 
$\beta$ may be more than $80^{\circ}$. 
Also stellar parameter somewhat differ, e.g. BM97 assume $\log L/L_{\odot}=2.9 \pm 0.2$, $M=4.8$~$M_{\odot}$, $R=5.1 \pm 1.1$ and $V_{\rm rot} =105$~km\,s$^{-1}$ whereas \cite{hub07a} give $\log L/L_{\odot}=2.3  \pm 0.1$,
$M=3.2 \pm 0.2$~$M_{\odot}$, $R=2.2 \pm 0.4$~$R_{\odot}$ and $V_{\rm rot} =$~46~km\,s$^{-1}$.
Since the adopted distance is the same, the stellar data and viewing geometry of IQ~Aur appear to be only moderately constrained 
and time-variability might play an additional role. The values
from \cite{koch06}, that are commonly used throughout the paper, are intermediate ones.

Both A0p stars are CP2 stars in the classification scheme from \cite{pres74} and might be expected to be overall similar given their identical spectral type.
However, IQ~Aur is more luminous, more massive and younger than $\alpha^{2}$\,CVn. 
Actually, IQ~Aur is extreme for an A0p star; as a main-sequence star it would be with its $B-V = -0.16$ of spectral type mid-B.
The adopted basic stellar data for IQ~Aur and $\alpha^{2}$\,CVn as compiled from \cite{boh93}, \cite{bych03}, \cite{koch06} and \cite{koch10} 
are summarized in Table\,\ref{log}.

\begin{table}[t!]
\setlength\tabcolsep{3pt}
\begin{center}
\caption{\label{log}Basic properties of our targets.}
\begin{tabular}{lll}
\hline\hline\\[-3mm]
Target & IQ Aur  &$\alpha^{2}$\,CVn \\
X-ray obs.& XMM & Chandra-ACIS-S \\
Dist. (pc) & 137 & 34\\
Sp.(chem) type & A0p - CP2, Si & A0p - CP2, Si\,{\tiny EuHgCr}\\
$\log L$\,($L_{\odot}$), $\log T$\,(K)& $2.5\pm 0.1$, $4.16 \pm 0.01$& $2.0 \pm 0.1$, $4.06 \pm 0.02$\\
$M$\,($M_{\odot}$), $R$\,($R_{\odot}$) & $4.0 \pm 0.1$, $2.9\pm 0.2$ & $3.0 \pm 0.1$, $2.5\pm 0.3$ \\
$\log t$ (yr), $\tau$ (frac.)& 7.8 (7.5-7.9),  0.4  &8.3 (8.1-8.4), 0.52 \\
$v\,sini$\,(km/s), $P_{\rm rot}$\,(d) & 56, 2.47 &18, 5.47 \\
$B_{\rm eff}$ (kG) &$ 0.74 \pm 0.4$ & $1.35 \pm 0.4$ \\
\hline
\end{tabular}
\end{center}
\end{table}

\section{Observations and data analysis}
\label{ana}

We observed $\alpha^{2}$\,CVn  in March 2009 with {\it Chandra} ACIS-S for 15~ks (Obs.ID 9923). 
The A0p star IQ~Aur was observed in
February 2010 with {\it XMM-Newton} for 115~ks (Obs.ID 0600320101) in the full frame mode with the thick filter inserted.
Due to the optical brightness of the target the OM (Optical Monitor) had to be closed.

We utilize the software packages CIAO\,4.2 and SAS\,10 to process and analyze the data, 
apply standard selection criteria and use the rather well 
calibrated energy ranges 0.2\,--\,8.0~keV ({\it XMM-Newton}) and 0.3\,--\,8.0~keV ({\it Chandra}).
Concerning the {\it XMM-Newton} observation of IQ~Aur, 
For the spectral analysis of the {\it XMM-Newton} EPIC (European Photon Imaging Camera) data of IQ~Aur
source photons are extracted from a 30\arcsec\, circular region around the source,
the background is taken from nearby source-free regions on the MOS and the PN detectors respectively. 
Of the EPIC detectors the PN provides a higher sensitivity, whereas the MOS provides a slightly better spatial and spectral resolution.
The signal in the RGS (Reflection Grating Spectrometer) detectors is rather weak, 
we therefore extracted high resolution RGS spectra from a 90\,\% PSF source region and
performed a cross-check with a narrower 70\,\% PSF extraction area, in which the background is further suppressed.
An additional problem are effective area drops in the RGS 
that affect e.g. the region around the \ion{O}{vii} intercombination line where several so-called 'bad' channels lead to a reduced number of detected counts.
To measure the line fluxes we use the fitting program CORA \citep{cora}, that accounts for these detector inhomogeneities.

We perform spectral analysis of the EPIC/ACIS-S data with XSPEC~V12.3 \citep{xspec}. To model the spectra we
use photo-absorbed, multi-temperature APEC models with abundances relative to solar values as given by \cite{grsa},
errors from spectral modeling denote 90\,\% confidence level. We note that the interdependence of
absorption, metallicity and emission measure ($EM = \int n_{e}n_{H}dV$) may result in different models of similar fit quality with the
consequence that absolute abundances are more poorly constrained than abundance ratios or relative changes.
Also the incomplete knowledge of atomic data used to calculate the spectral models adds to these uncertainties, however they do not
alter the main conclusions of this work.

\section{Results}
\label{res}
Here we report on the results relevant to the origin of X-ray emission in Ap/Bp stars and present
an analysis of the IQ~Aur data, model its spectra, take a detailed look at the flare and investigate X-ray lines from the high-resolution spectrum.

\begin{figure}[t]
\hspace*{0.3cm}
\includegraphics[height=45mm]{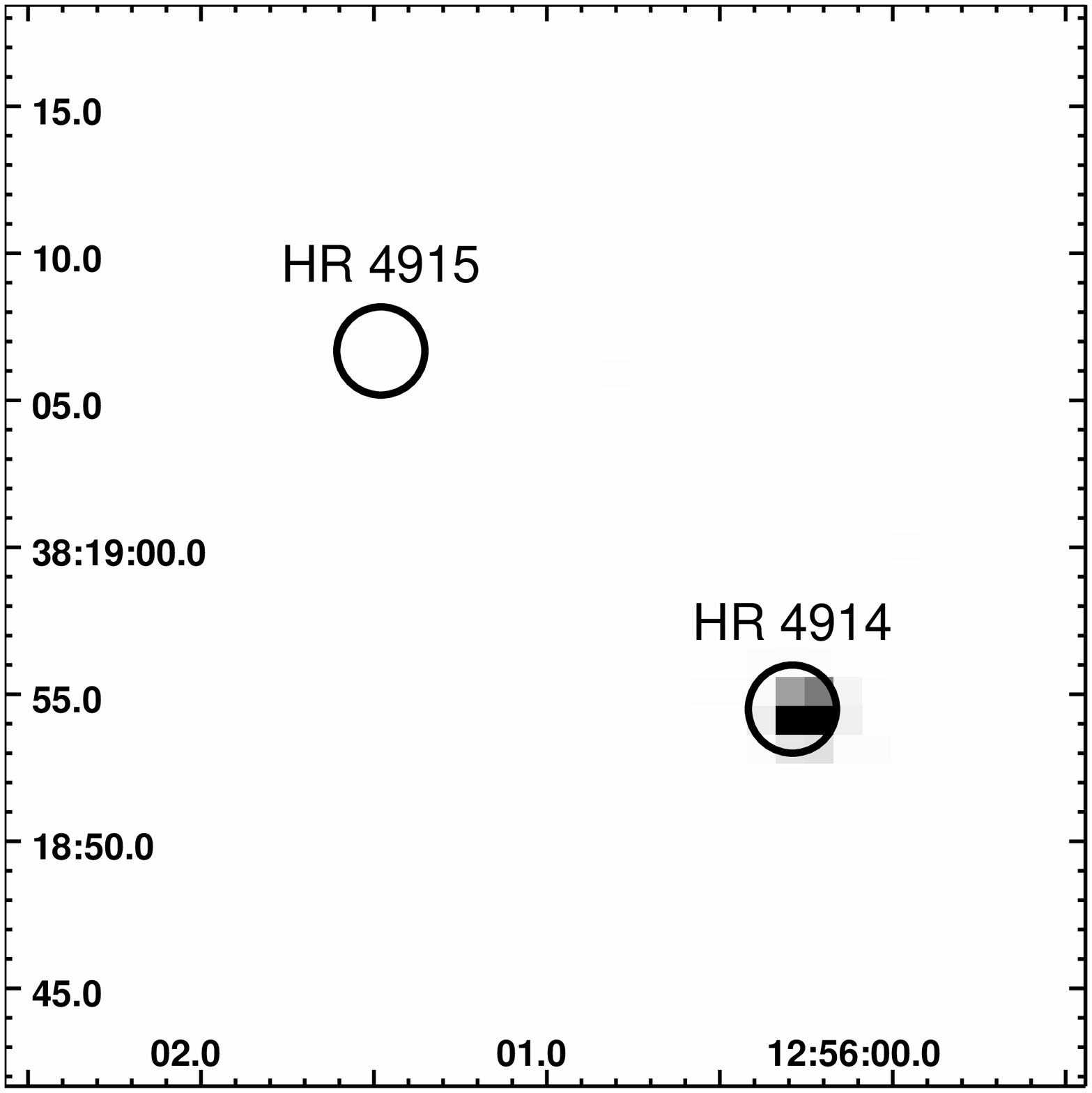}
\hspace{5mm}
\includegraphics[height=45mm]{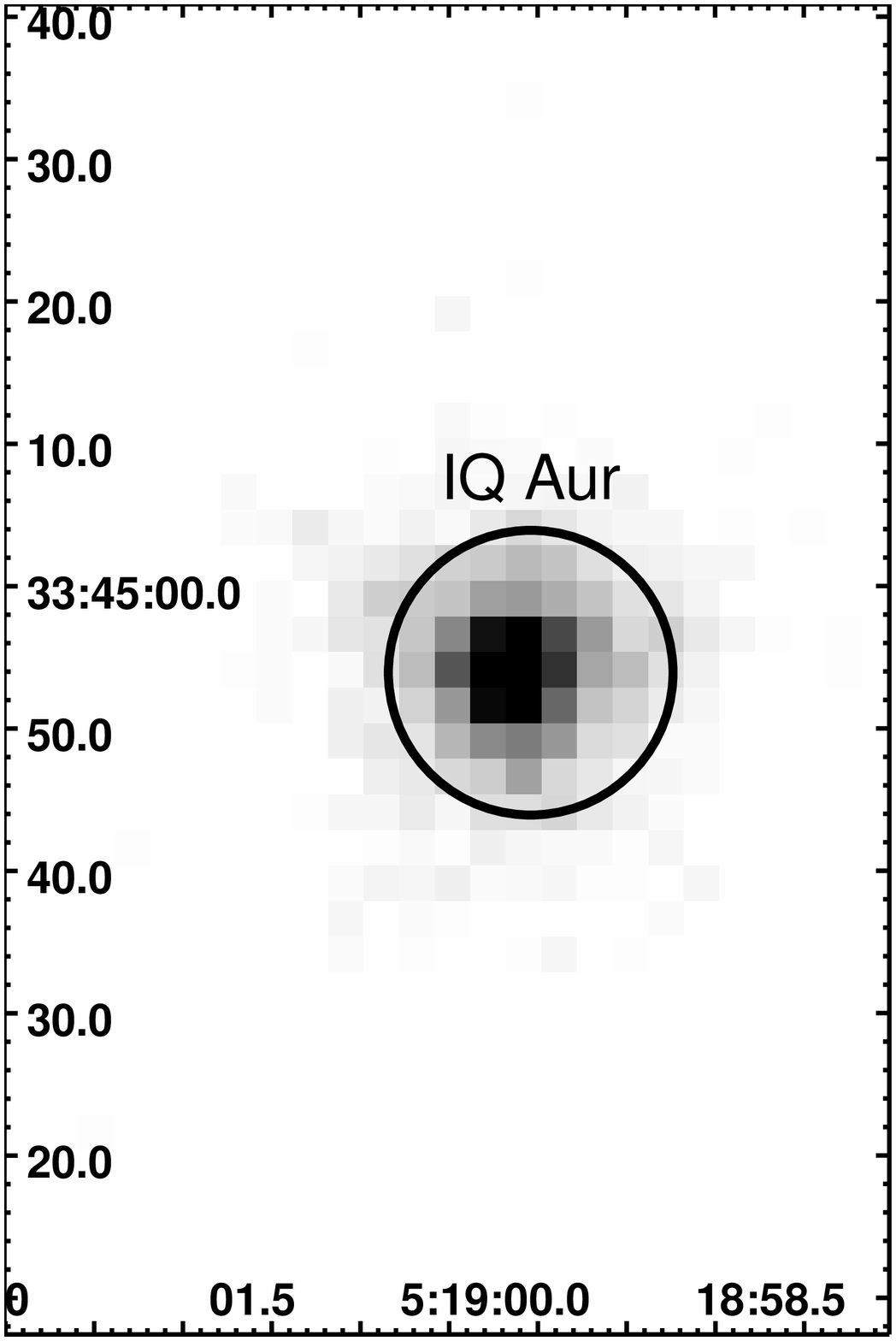}
\caption{\label{im}{\it Left panel:} ACIS-S image of the $\alpha^{2}$\,CVn, the A0p star HR~4915 is X-ray dark. {\it Right panel:} MOS1 image of the A0p star IQ Aur.
The circles denote the respective optical positions.}
\end{figure}

\subsection{Source detections, light curves and a flare}
\label{sl}

The ACIS-S image of the $\alpha^{2}$\,CVn field, shown in Fig.\,\ref{im}, clearly reveals that 
the A0p star $\alpha^{2}$\,CVn (HR~4915) is X-ray dark. In the 0.3\,--\,3.0~keV band we detect no photon in a 2\,\arcsec source region (95\,\% PSF) 
centered on the expected optical position at an estimated background
of 0.2~photons. We detect the F0 star HR~4914 with an X-ray luminosity of $\log L_{\rm X}= 28.6$~erg\,s$^{-1}$, a rather typical value.
For its position we find a small offset of about 0.5\,\arcsec~between the X-ray and expected optical position;
nevertheless, applying a spatial correction to HR~4915 does not change the outcome.
Adopting Poissonian statistics we derive an 95\,\% conf. upper limit of three counts for the source region. 
Assuming a plasma temperature of 0.3\,(0.8)~keV this corresponds to 
an upper limit on the X-ray luminosity of $L_{\rm X} < 0.6~(1.0) \times 10^{26}$~erg\,s$^{-1}$,
three to four orders of magnitude below the X-ray luminosity of IQ~Aur ($L_{\rm X}= 4 \times 10^{29}$~erg\,s$^{-1}$), despite both being apparently similar stars.
The flux limit for $\alpha^{2}$\,CVn converts into the by far the lowest upper limit obtained on the X-ray brightness for an A0p star.

In contrast, IQ~Aur is detected in all {\it XMM-Newton} detectors; see the right panel of Fig.\,\ref{im}, where we show the MOS1 image.
The derived X-ray position is about $\lesssim 1$\,\arcsec~off from the expected optical position, a value within the positional uncertainty for {\it XMM-Newton}.
To investigate the X-ray variability of IQ~Aur,
we extracted photons from a 35\,\arcsec region ($\sim$ 90\,\% encircled energy) around the X-ray source position on the EPIC detectors.

\begin{figure}[t]
\includegraphics[height=62mm,width=90mm]{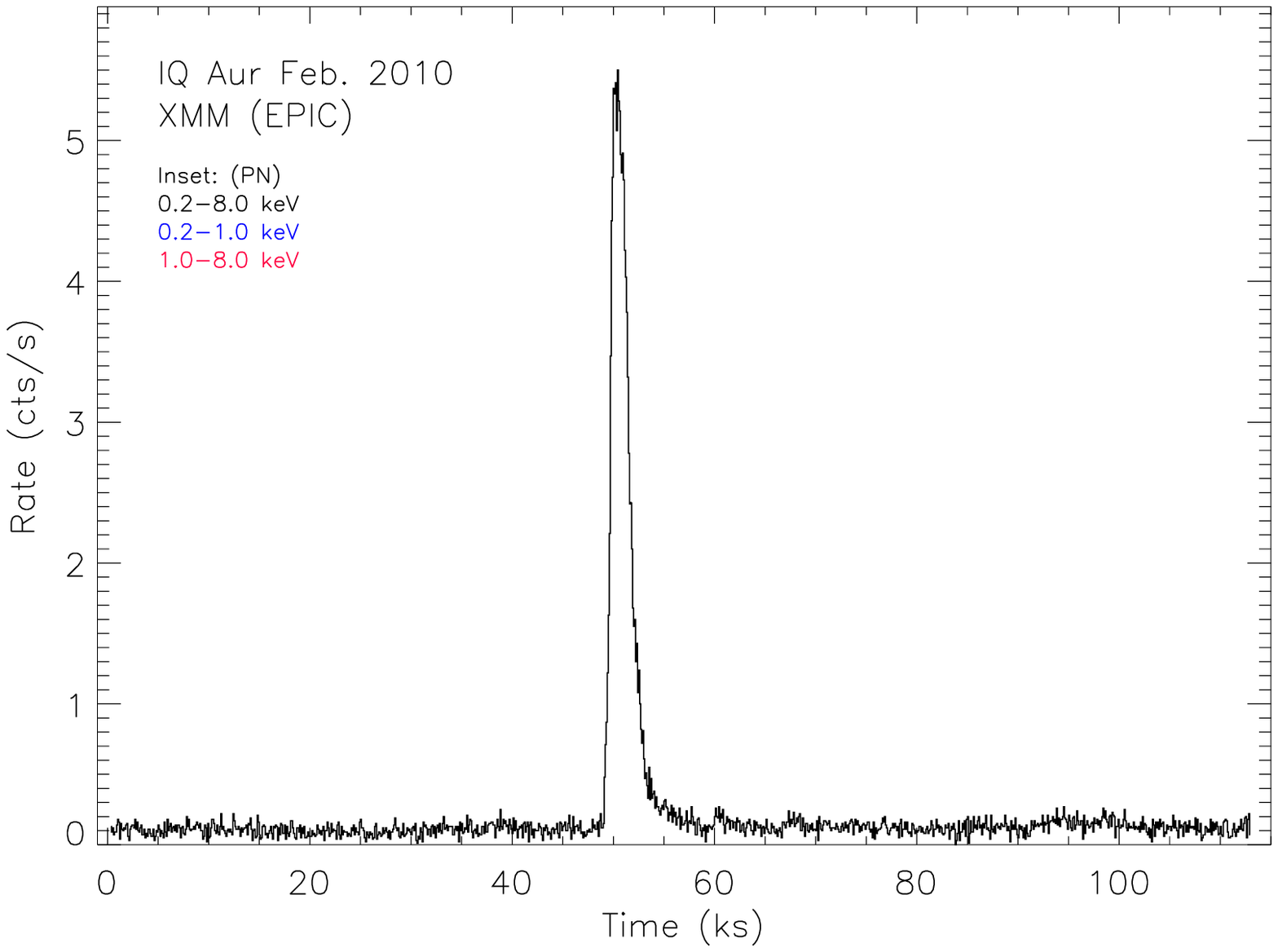}

\vspace*{-5.7cm}
\hspace*{5.5cm}
\includegraphics[height=45mm]{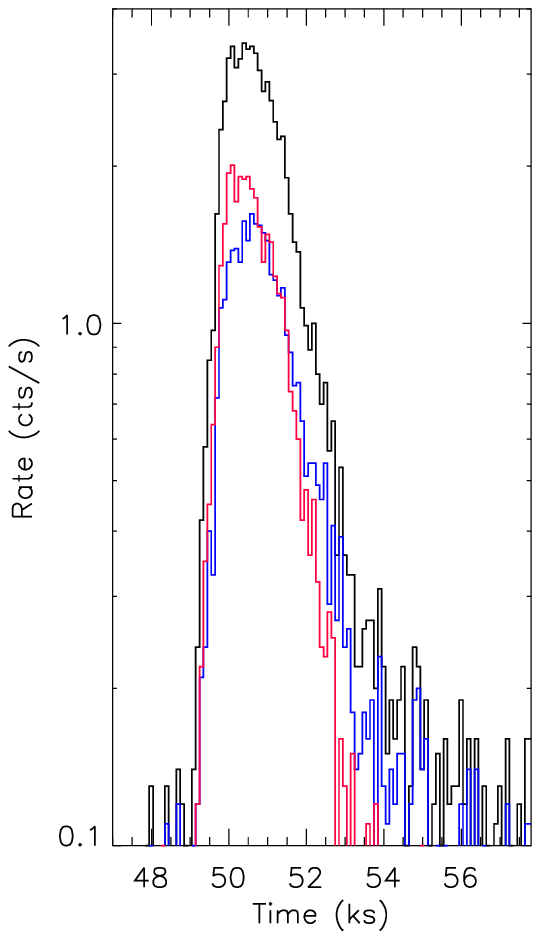}
\vspace*{1cm}
\caption{\label{lc}X-ray light curve of IQ Aur from the summed EPIC data in the 0.2\,--\,8.0~keV band. {\it Inset:} The flare on a logarithmic scale in three energy bands as seen by the PN detector.}
\end{figure}

\begin{figure}[t]
\includegraphics[width=90mm]{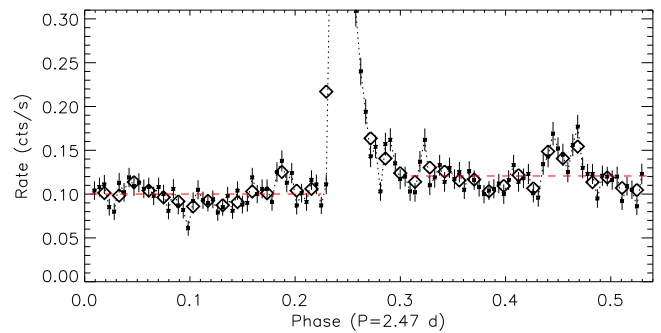}
\caption{\label{lcqq}Quasi-quiescent part of the X-ray light curve, phased with the rotation period of IQ Aur and binned to 1~ks (asterisks) and 3 ks (diamonds).
The average rate before and after the flare is indicated by a red dashed line respectively.}
\end{figure}

In Fig.\,\ref{lc} we show the background subtracted light curve with 100~s time binning. 
Beside minor variability that is present throughout the observation,
the X-ray count rate increases after 49~ks by a factor of roughly 40 within less than half an hour,
the total flare event has a duration of about two hours.
In the inset of Fig.\,\ref{lc} we show the light curve of the flare part as seen by the PN detector, also separately in a soft and hard energy band.
The shape of the light curve with its fast rise time of 1000~s and e-folding decay time of $\sim$\,1200\,--\,1400~s points to a single, rather compact flaring structure.
Since for X-ray emitting structures/loops 
the radiative energy loss scales with the square of the particle density,
compact, i.e. small and dense structures, 
radiate their energy much faster away than larger tenuous structures of comparable X-ray luminosity.
Further, the harder X-ray
emission increases by a larger factor and its peak precedes the one of the softer emission,
a behavior that is reminiscent of coronal flares.
We do not detect any spatial shift between the photons detected during the quasi-quiescent and the flaring phase at an accuracy of $\lesssim 0.5$\arcsec.
In Fig.\,\ref{lcqq} we show the quasi-quiescent part with two different time binnings.
Notably, the average quasi-quiescent X-ray count rate has increased by roughly 20\,\% higher after the flare and the light curve shows slightly stronger variability, although
the spectral hardness given by the ratio of the count rates in the 0.2\,--\,0.8~keV vs. 0.8\,--\,2.0~keV energy band 
is virtually identical compared to the pre-flare phase.

We do not see a strong modulation of the X-ray light curve that might be expected in an oblique rotator model ($P_{\rm rot}= 213$~ks).
Significant, roughly sinusoidal shaped rotational modulation would be present, if the star partially occults
X-ray emitting material depending on rotational phase or when a disk around the star is viewed under different angles. An example are the X-ray brightness
variations of the massive magnetic star $\theta ^1$~Ori\,C \citep{gag05}.
This model assumes cylindrical symmetry for the system as is likely appropriate for a star with a disk 
and what is commonly observed when e.g. magnetic field variations are monitored, 
but clearly we cannot completely exclude unexpected phenomena during the not covered phase.  
Brightness variations on timescales of 20~ks ($\sim$~0.1 phase) between consecutive bins during the pre-flare 
or the post-flare phase have an amplitude of $\lesssim 10$\,\% and appear rather irregular.
However, the strength of the modulation depends on the plasma location and the geometry of the system 
and for a small obliquity the effects are likely minor in any case.

\subsection{Spectral properties in quasi-quiescence}
\label{ss}

\begin{figure}[t!]
\includegraphics[height=88mm,angle=-90]{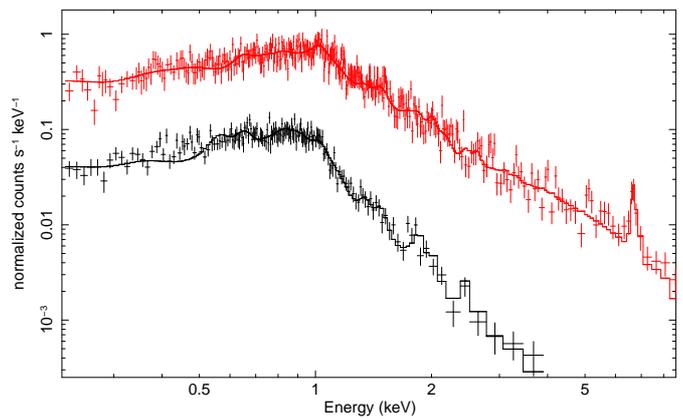}
\caption{\label{spec}XMM/PN spectra of the quasi-quiescent (black) and the flaring (red) phase with applied spectral models.}
\end{figure}

The global spectral properties of IQ~Aur are determined from the EPIC spectra
by using photo-absorbed multi-temperature models with free metallicity.
Given the large variations in count rate over the observation, we separate the data into a flaring phase (PN: 49.1\,--\,58.1 ks) and
a quasi-quiescent (QQ) phase for spectra analysis. As an example we show the respective PN spectra with the
applied spectral models in Fig.\,\ref{spec}; the derived spectral properties from our modeling are summarized in Table\,\ref{fit}.

The description of the quasi-quiescent spectrum clearly requires, beside the 0.3~keV component seen by {\it ROSAT}, 
the presence of additional hotter plasma components.
Applying a one temperature model for example to the MOS spectra results in an average temperature of $kT=0.75$~keV, albeit with $\chi^2_{\rm red}>2$ 
the fit is quite poor. Models with three temperature components describe the individual data sets rather well. 
The absorption component required to model the X-ray spectra of IQ~Aur is rather weak, compatible with
the findings from the {\it ROSAT} data. 
We again find strong emission from rather cool plasma at $\sim3$~MK, but overall the plasma temperatures derived from the {\it XMM-Newton}
spectra are significantly higher than those determined with {\it ROSAT}, where only weak indications for hotter plasma were found due to the limited
sensitivity of the instrument. 
We detect a significant contribution from $\gtrsim10$~MK plasma even in the quasi-quiescent phase of IQ~Aur, here it accounts for about 30\,\% of the total emission measure.
We find that a subsolar metallicity best describes the data.
Using solar abundances reduces the emission measure, but also worsens the model quality ($\chi^2_{red}=1.24$),
albeit the temperature structure remains nearly identical.
Allowing individual abundances to vary independently does not lead to robust results, therefore a global metallicity is used in all models.
The best fit model corresponds to a source flux of $1.7 \times 10^{-13}$~erg\,cm$^{-2}$\,s$^{-1}$
and an X-ray luminosity of $L_{\rm X}= 4.2 \times 10^{29}$~erg\,s$^{-1}$ 
in the 0.2\,--\,2.0\,~keV band; for the {\it ROSAT} 0.1\,--\,2.4~keV band we derive a roughly 20\,\% higher flux.
This corresponds to an activity level of $\log L_{\rm X}$/$L_{\rm bol} \approx -6.4$, 
assuming that the X-rays are emitted by IQ~Aur. Considering the {\it ROSAT} detection at a very similar X-ray luminosity of
$L_{\rm X}= 4.0 \times 10^{29}$~erg\,s$^{-1}$ nearly 15~years ago, IQ~Aur is likely
emitting X-rays at a rather constant level for more than a decade.

\subsection{Global flare properties}

During the flare phase, IQ~Aur has not only significantly brightened, but as shown in Fig.\,\ref{spec}, also the associated spectrum is much harder.
The flare is most prominent at high temperatures, but large excess emission is also 
present at lower temperatures.
During the peak of the strong flare the emitted X-ray luminosity of IQ~Aur rises to about $L_{\rm X}= 3.2 \times 10^{31}$~erg\,s$^{-1}$ (0.2\,--\,10.0~keV),
corresponding to a flux increase of nearly a factor 100 compared to the quasi-quiescent state. 
With an average excess luminosity of $L_{\rm X}= 6.2 \times 10^{30}$~erg\,s$^{-1}$ and a duration of 9000~s, 
the flare released about $E_{\rm X}= 5.6 \times 10^{34}$~erg only at soft X-ray energies. 
The flare emission is dominated by extremely hot plasma and even the 
average temperature is found to be around 45~MK.
In the PN spectrum the prominent emission line complex from He-like \ion{Fe}{xxv} is clearly visible at 6.7~keV, consistent with 
a hot plasma component at temperatures of around 60\,--\,70~MK as deduced from the spectral model. 
Weak excess emission is present at the low energy tail of this feature,
that could be due to fluorescent emission from neutral iron from the Fe\,K$\alpha$ line at 6.4~keV. However, formally the line is not detected at the 90\,\% CL and its
contribution to the X-ray flux is with  an equivalent width of EW\,$\lesssim 100$~eV at best marginal. 
No significant changes in the absorption component are detected;
in contrast, the metallicity of the flare plasma is found to be significantly higher than those of the quasi-quiescent phase.
While the absolute scale is rather poorly constrained, 
we find a distinct metallicity trend, most noticeable when further separating the flare into individual time intervals as discussed in Sect\,\ref{sf}.

\begin{table} [t!]
\begin{center}
\caption{\label{fit} Spectral models of IQ Aur (EPIC data).}
\begin{tabular}{lrrl}
\hline\hline\\[-3mm]
Par.  & QQ& Flare & Unit\\\hline\\[-3mm]
$N_{\rm H}$& 1.1$\pm 0.9$ & 1.0$\pm$0.9& $10^{20}$cm$^{-2}$\\
Abund.& 0.31$\pm$0.08 & 0.67$\pm$0.28  & solar\\
$T_{1}$& 0.26$\pm$0.03 & 0.71$\pm$0.09 & keV\\
$EM_{1}$&1.75$\pm 0.71$ & 4.7$\pm2.0$ & $10^{52}$cm$^{-3}$ \\
$T_{2}$ &0.64$\pm$0.15 & 1.59$\pm$0.19 & keV\\
$EM_{2}$& 1.58$\pm0.56$& 14.4$\pm6.9$  & $10^{52}$cm$^{-3}$\\
$T_{3}$&1.10$\pm$0.11 & 6.18$\pm1.46$ & keV\\
$EM_{3}$& 1.88$\pm0.66$&  22.4$\pm7.1$ & $10^{52}$cm$^{-3}$\\\hline\\[-3mm]
$\chi^2_{\rm red}${\tiny(d.o.f.)}& 1.09 (413) & 0.97 (555)&\\\hline\\[-3mm]
$L_{\rm X}${\tiny(em.\,0.2-10.0\,keV)}& $4.5 \times 10^{29}$ & $6.7 \times 10^{30}$& erg~s$^{-1}$\\\hline
\end{tabular}
\end{center}
\end{table}

\subsection{The flare under scrutiny}
\label{sf}

\begin{figure}[t]
\includegraphics[width=88mm]{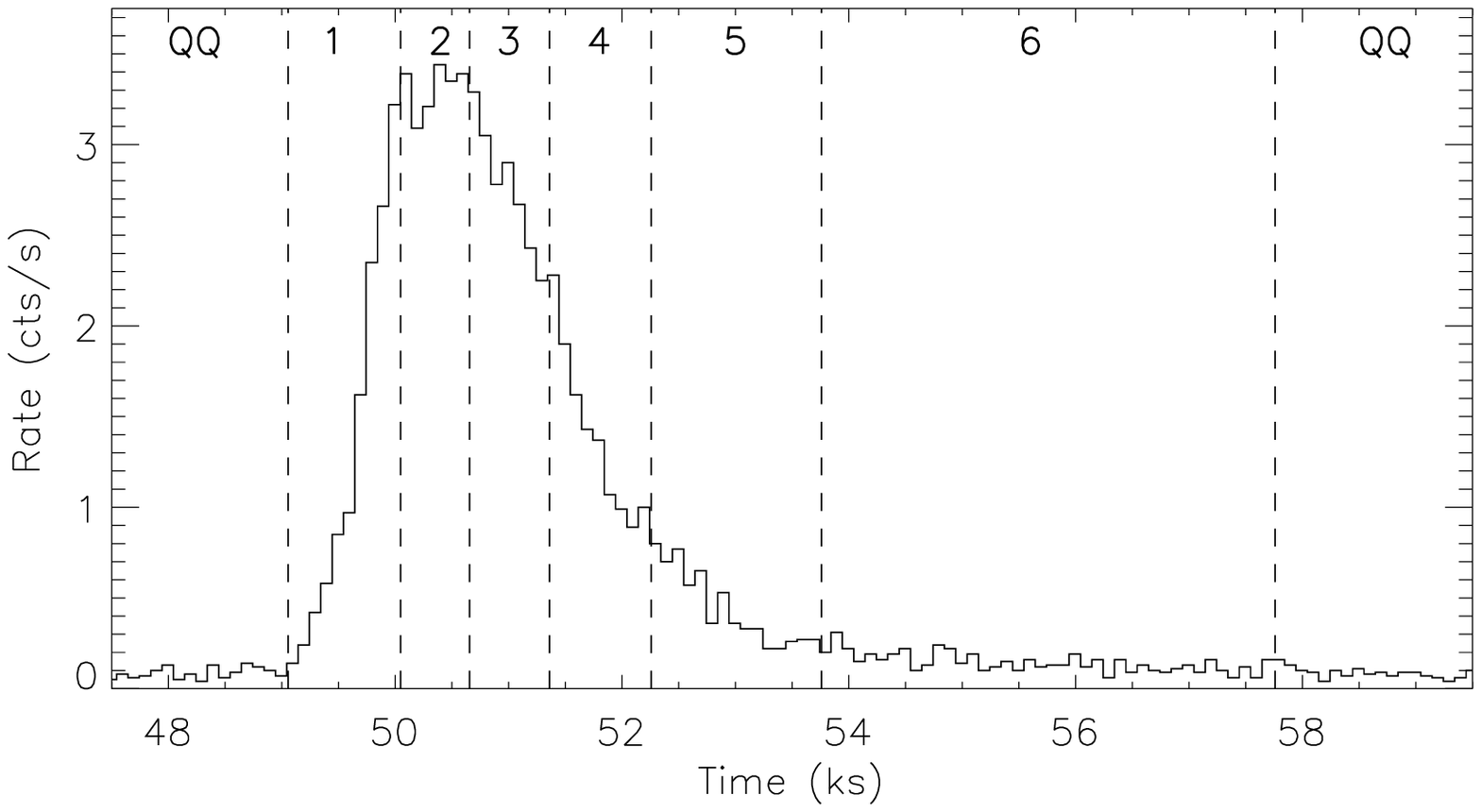}
\includegraphics[width=88mm]{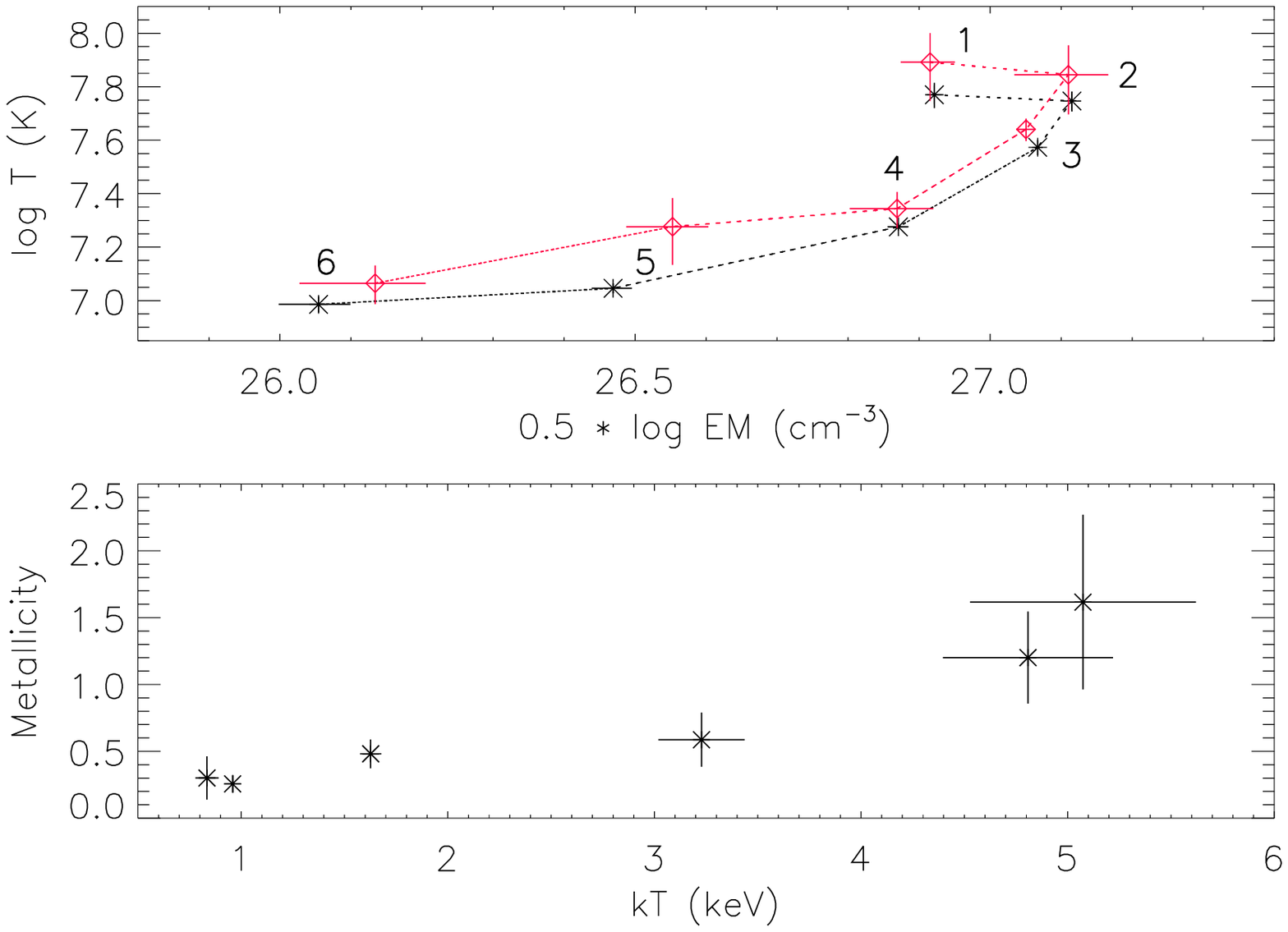}
\caption{\label{flp}Time resolved study of the IQ~Aur flare with XMM/PN data. {\it Top:} time intervals used in the analysis. 
{\it Middle:} temporal evolution of the plasma in the temperature/density plane (black/red: 1/2-T model).
{\it Bottom:} metallicity vs. plasma temperature.}
\end{figure}

To study the evolution of the flare and its underlying structure in greater detail, we perform a time resolved analysis
for each of the six time segments as shown in the upper panel of Fig.\,\ref{flp}. To derive the properties of the flare plasma (1-6), we added
plasma components to the quasi-quiescent (QQ) model derived for the PN data. Here we use models with one or two temperature components and with or without metallicity
as a free parameter to describe the flare plasma. The models lead to overall consistent results, see for example the
time-evolution of the plasma properties presented in the middle panel of Fig.\,\ref{flp} for models with solar metallicity. 
The temperature given for the 2-T model is the emission measure weighted temperature.
The flare reaches around maximum an X-ray brightness of about $\log L_{\rm X}=31.5$~erg\,s$^{-1}$ and an average plasma temperature of around 80~MK, 
shortly after the impulsive heating during the rise phase (bin 1), even roughly 100~MK plasma is present. The maximum temperature leads the maximum
emission measure and thus in general the flare shows the temporal evolution of 
a typical coronal flare, i.e. a clock-wise turn in the temperature vs. density plane \citep[see e.g.][]{gue04}.
Here the square-root of the emission measure is used as a density proxy, thus
a fixed spatial extend of the X-ray emission region is assumed.
While the initial decay is apparently steep, but only poorly constrained (bins 2-3), the flare evolution trajectory shows some 
flattening (bins 3-6) of the decay path.
The origin of this behavior is unknown; it could be due to minor re-heating events in the same structure or produced by additional reconnection events
in other magnetic structures, that again might be triggered by the large flare. 
Another possibility is a change in size of the emitting region over the duration of the event.

Assuming a single, loop-like emitting structure we can derive its size by adopting the formalism described in \cite{rea97} 
and parameters applicable to XMM/EPIC as given in \cite{rea07}. 
By using the maximum plasma temperature and the flare decay time obtained from the light curve one can determine
the size of the flaring structure, that is treated as an instantaneously heated coronal loop. 
Basically one finds that $L\propto \tau_{lc} \sqrt{T_{\rm max}}$, with $L$ being the loop half-length, $T_{\rm max}$ the maximum plasma temperature and $\tau$ the
decay time from the X-ray light curve and an additional scaling factor. This factor depends on the amount of additional heating during the decay, thus
one needs to estimate the importance of sustained heating. This is done by fitting 
the decay path derived from the one temperature model with a straight line to
about about one tenth of the maximum intensity. This corresponds to bins 2\,--\,5 in the middle panel of Fig.\,\ref{flp}, where we derive a slope of $\sim 1$ for the decay
and thus moderate re-heating has to be taken into account.
Adopting the appropriate modeling parameter, i.e. $T_{\rm obs} =65$~MK, an e-folding decay time of $\tau = 1200$~s (measured to 1/10 of the maximum) and a correction factor for
a slope of one, we obtain $L = 1.8 \times 10^{10}$~cm ($0.1~R_{*}$) for the loop half-length.

Since a coronal loop model might not be perfectly appropriate for IQ~Aur and possible heating at later phases of the flare is only poorly constrained,
we use as a conservative approach our most extreme values, i.e. 100\,MK plasma temperature and 1400\,s decay time and ignore ongoing heating. 
This method results in an upper limit for the size and we
derive $L \lesssim 3.5 \times 10^{10}$~cm ($0.2~R_{*}$) as maximum extent of the structure. While the flaring structure might be much smaller, even 
the upper limit corresponds to a moderately sized structure when compared to the stellar dimensions of IQ~Aur.
This finding excludes very large magnetic structures, e.g. from the global dipole,
as origin of the flare.

The IQ~Aur flare is accompanied by a significant change in the metallicity of the X-ray emitting plasma as illustrated
in the lower panel of Fig.\,\ref{flp} where we plot the results obtained from a 1-T model with metallicity as free parameter.
During the impulsive phase of the flare as traced by the plasma temperature, more metal rich material is heated to X-ray temperatures.
The freshly heated plasma from the large flare exhibits solar or even slightly super-solar metallicity and thus
a two to three times higher metallicity than the X-ray emitting plasma observed in the quasi-quiescent phase.
While the absolute scale is only moderately constrained as outlined in Sect,\,\ref{ana}, the relative changes are quite robust and also found in two temperature models or the analysis the MOS spectra.
In the final stages of the flare evolution the remaining material has cooled down
to temperatures below 1~keV, exhibits pre-flare metallicity and becomes rather indistinguishable from the quasi-quiescent plasma.
Overall, we find a clear correlation between plasma temperature and metallicity, indicating a strong modification of the chemical composition during the event.
Notably, the derived characteristics are quite typical for a stellar flare \citep[see e.g.][]{gue04}.

\subsection{The high resolution X-ray spectrum}
\label{hires}

\begin{figure}[t]
\includegraphics[width=88mm]{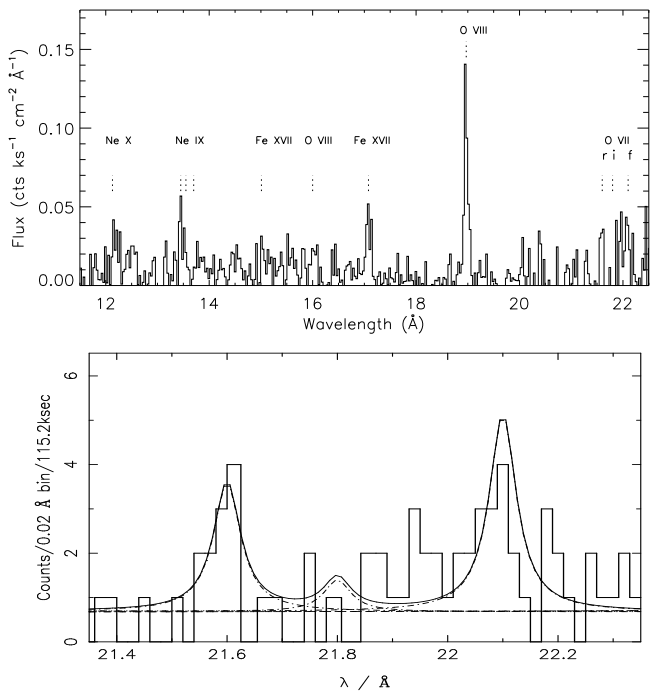}
\hspace*{0.2cm}
\includegraphics[width=45mm,height=80mm,angle=-90]{16843f6b.ps}
\caption{\label{rgs}High resolution X-ray spectrum of IQ Aur. {\it Top:} Total RGS\,1+2 merged and flux-converted. 
{\it Bottom:} \ion{O}{vii} triplet with model accounting for effective area inhomogeneities.}
\end{figure}

The high resolution RGS spectrum from IQ~Aur has unfortunately only moderate signal to noise and only rather strong lines were clearly detected even in the 
total spectrum; see Fig.\,\ref{rgs}, where we show a flux-converted spectrum of the merged RGS detectors from a 90\,\% PSF extraction region. 
The individual line fluxes are measured from the total count spectrum and
cross-checked by using a broad and a narrower 70\,\% PSF extraction region.
The lower panel of Fig.\,\ref{rgs} shows the \ion{O}{vii} triplet with applied model, confirming that
the forbidden line (22.1 \AA) is significantly stronger than the intercombination line (21.8 \AA). 
The S/N ratio is quite poor, even in the total spectrum that includes the flare.
We also investigated the quasi-quiescent and flare phase separately, but except stronger emission from 'hotter lines' (e.g. \ion{Fe}{xvii}) during the flare,
a quantitative comparison suffers from the low S/N.
The fluxes of emission lines used in the subsequent analysis as determined from the total spectrum are given in Table~\ref{linetab}.
They are corrected for an absorbing column of $\log N_{\rm H}=20$~cm$^{-2}$ and
we expect a flare contribution of roughly 30\,\% to the given line fluxes.

\begin{table}[!ht]
\caption{\label{linetab}Oxygen lines used in the analysis.}
\begin{center}{
\begin{tabular}{lcc}\hline\hline
Ion/Line& Wavelength  & Flux \\
 & (\AA) & ($10^{-15}$ erg cm$^{-2}$~s$^{-1}$)\\\hline\\[-3mm]
\ion{O}{viii} & 18.97 & 14.7$\pm$2.3 \\
\ion{O}{vii}(r) & 21.60 & 4.5$\pm$1.9 \\
\ion{O}{vii}(i) & 21.80 & 1.6$\pm$1.2 \\
\ion{O}{vii}(f) & 22.10 & 5.9$\pm$1.9\\\hline
\end{tabular}}
\end{center}
\end{table}

We utilize the He-like \ion{O}{vii} triplet to
determine the distance of the X-ray emitting plasma to the surface of IQ~Aur.
\ion{O}{vii} traces the location of plasma at temperatures around 2~MK
and its forbidden ($f$) to intercombination ($i$) line ratio is sensitive to density and UV radiation, specifically one finds
\hbox{$f/i=R_{0}/(1+\phi/\phi_{c}+n_{e}/n_{c})$}, see e.g. \cite{mewe78,por01}.
In the low density case ($n_{e}\lesssim 10^{10}$~cm$^{-3}$) additional photons in the $i$-line are only due to the radiation term $\phi/\phi_{c}$.
The radiation term describes the strength of the UV flux at 1630\,\AA\, and thus
depends on the effective temperature and distance to the star.
Using the fluxes in Table~\ref{linetab} we obtain $f/i=3.7\pm 3.0$, unfortunately the errors are relatively large.
The ratio is consistent with virtually no radiation field being present ($R_{0}=3.95$), but much lower values are formally also allowed. 
Adopting $f/i\gtrsim 1$, a photosphere with $T_{\rm eff}=14450$~K and evaluating the radiation term, one finds that 
the X-ray emitting region is either shielded against the stellar UV-field or
the X-rays originate predominantly from $\gtrsim 7~R_{*}$ above the surface of IQ~Aur. 
As an alternative approach, we use the UV flux as measures by the IUE satellite and calculate $T_{\rm rad}$ from the observed spectrum.
The 1630\,\AA\, UV flux from IQ~Aur varies by a factor of nearly three over several years;
using the lower flux measurements we find that the X-ray emitting region has to be located above the star at distances of $d \gtrsim 5~R_{*}$. 
This result is independent of our assumptions; if higher densities are present, even larger distances to the stellar surface are required.
Similarly, applying a correction
for the weak extinction towards IQ~Aur or adopting a higher UV flux would again shift the X-ray plasma location to larger distances.
Therefore we can virtually exclude X-ray emitting structure on or close to the star and
thus the X-rays are produced at least several stellar radii above the surface of IQ~Aur.

\begin{figure}[t]
\includegraphics[width=90mm]{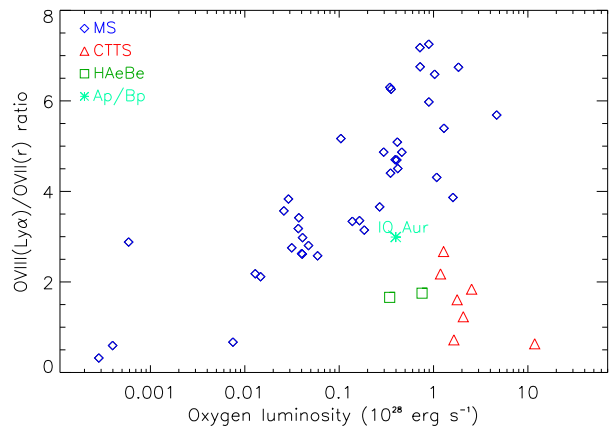}
\caption{\label{coolex}Show above is the quasi-quiescent \ion{O}{viii}/\ion{O}{vii} ratio of IQ Aur in comparison to accreting/wind driving young stars 
and main sequence stars.}
\end{figure}

We also investigate the quasi-quiescent and flare phase alone and find for both cases a similar large $f/i$-ratio, however with substantial error.
Similarly, we study the \ion{Ne}{ix}-triplet and find an $f/i$ ratio that is slightly higher but again fully
consistent with the low density/no UV-field limit, however no additional information is gained to \ion{O}{vii}.
Further, we use the density constraint derived from
the \ion{O}{vii} $f/i$-ratio, as a very rough cross-check on the size of the emitting volume. 
Adopting $f/i\gtrsim 1$ implies a density of 
$n_{e} \lesssim 5 \times 10^{10}$~cm$^{-3}$, therefore from this value we can now formally derive a minimum size of the emission region. 
The average flare EM is $4\times 10^{53}$~cm$^{-3}$ with $EM \approx 0.85~n_{e}^{2} V$ and thus one obtains for an emitting sphere 
a radius of $r\approx 0.14~R_{*}$. We again find a rather small structure, 
in agreement with the estimations $L < 0.2~R_{*}$ derived from the above analysis of the flare.

By using temperature sensitive \ion{O}{viii}/\ion{O}{vii} line ratios, we test if additional cool plasma
compared to pure coronal sources is produced by IQ~Aur (see Fig.\,\ref{coolex}).
The \ion{O}{vii} lines form predominantly at temperatures around 2~MK, while \ion{O}{viii} traces
hotter plasma at 3\,--\,5~MK.
Additional \ion{O}{vii} emission may be produced by accretion or wind shocks, as commonly observed in young T~Tauri and Herbig~AeBe stars,
that generate rather cool plasma compared to coronal sources
of comparable X-ray brightness, see e.g. \cite{rob07b}.
The \ion{O}{viii}/\ion{O}{vii} ratio measured for the quasi-quiescent state IQ~Aur is slightly below the values found for coronal sources, 
i.e. its X-ray emission is only moderately soft for its X-ray luminosity.
However, it is well above the ratios found for HAeBe stars; young intermediate mass stars that drive outflows 
and jets with velocities of a few hundred km/s. These stars
produce X-ray emission from predominantly cooler plasma, likely related to shocks in their outflows \citep{guenther09}.

\section{Discussion}
\label{dis}
Our X-ray observations of the A0p stars IQ~Aur and $\alpha^{2}$\,CVn present quite a complex X-ray picture.
Here we discuss some implications imposed by our findings in the framework of the magnetically channelled wind shock (MCWS) scenario vs. a possible low-mass companion.

\subsection{MCWS models for Ap/Bp stars}

To detect MCWS generated X-ray emission from A0p stars, some basic conditions have to be fulfilled. 
The shock-speed has to be large enough to heat the plasma to X-ray emitting temperatures, the mass loss rate has be high enough to produce
significant amounts of hot plasma to be detected with current instruments and a magnetic field configuration has to be present that effectively
supports the conversion of kinetic wind energy to X-ray emission.

A key parameter in the dynamic MCWS model is the magnetic confinement parameter $\eta_{*}$ \citep{dou02}.
It basically describes the ratio between the energy density of the magnetic field and the kinetic wind energy density
and depends on stellar properties as given by $\eta_{*} \propto B^{2}_{\rm eq}R^{2}_{*}/\dot{M}V_{\infty}$; here $B_{\rm eq}$ is the equatorial field strength,
$R_{*}$ the stellar radius, $\dot{M}$ the mass loss rate and $V_{\infty}$ the terminal wind speed.
Adopting values for IQ~Aur, $B_{\rm eq} =2$~kG, $R_{*}=2.9~R_{\odot}$, $\dot M=10^{-11}~M_{\odot}$\,yr$^{-1}$ and $V_{\infty}=700$~km\,s$^{-1}$,
gives a confinement parameter in the order of $\eta_{*} \sim 10^{6}$. This value has  significant uncertainty, but
definitely the requirements for strong confinement, i.e. $\eta_{*} \gg 1$, are fulfilled. 
While at first analytic studies of rigidly rotating magnetospheres
indicated that in the case of strong magnetic confinement an equatorial disk builds up from corotation radius outwards \citep{tow05},
its formation is also present in MHD simulation when including effects of stellar rotation. 
These models predict the formation of a rigidly rotating disk by the accumulation of wind material, however
episodes of infall and breakouts limit the build-up of the disk and lead to a dynamic but quasi-stationary behavior \citep{dou08}; moreover
the location of the temporary confined plasma around the equatorial plane can be estimated with the formalism presented in these works.
For moderate rotation and very strong confinement the inner edge of disk is roughly given by the 'associated Kepler-radius' 
$R_{\rm K}= W^{-2/3}$ with $W=V_{\rm rot}/V_{\rm orb}$ and $V_{\rm orb}=\sqrt{GM/R_{*}}$; using $V_{\rm rot}=50$~km\,s$^{-1}$ we obtain $R_{\rm K} \sim 5 R_{*}$.
The outer edge is determined by the extent of the closed magnetosphere $R_{\rm C} \approx 0.7 R_{\rm A}$, slightly below the 
Alfv{\'e}n radius given by $R_{\rm A} \approx \eta_{*}^{1/4} \times R_{*}$. 
For IQ~Aur we obtain $R_{A} \approx 30~R_{*}$; clearly we are in a regime where $R_{\rm A} > R_{\rm K}$ and thus the conditions required by the model
for the formation of a rigid disk are fulfilled.
The disk of IQ~Aur is expected to be roughly located between
$\approx 5\,-\,20~R_{*}$ and breakout events are mainly launched at distances of $\approx 20\,-\,30~R_{*}$.
We caution that these models assume an aligned dipole and were derived for more massive stars,
thus an extrapolation to our Ap/Bp stars may not be straightforward.
Nevertheless, the model is generally applicable for magnetically channelled line-drive stellar winds and thus appears sufficient for the
discussion of the overall phenomenology of X-ray emission presented in the following.

\subsection{Why is IQ~Aur X-ray bright and $\alpha^{2}$\,CVn X-ray dark?}

The X-ray luminosity produced by MCWS follows the relation $L_{\rm X}\propto \dot M V_{\infty} B^{0.4}_{*}$, whereas the plasma temperature is given by
$T_{\rm X} \approx 1.15 \times10^{5} {\rm K}\,(V_{\rm sh}/100~\rm{km\,s}^{-1})^{2}$ (BM97).
Adopting similar wind speed and mass loss rate, 
one expects $\alpha^2$\,CVn to be comparably bright or even moderately X-ray brighter than IQ~Aur.
However, it is at least a factor thousand X-ray fainter than 'expected' from the simplified assumption. 
This implies that either the mass loss rates differ by orders of magnitudes or that shock-speeds for $\alpha^{2}$\,CVn are so low, 
that the plasma does not reach X-ray temperatures. Since stellar wind speeds are likely comparable and strong confinement is achieved for both stars, a very
low plasma temperature of $T_{\rm X} \lesssim 3 \times 10^{5}$~K as
would be required to explain the tight upper limit on $L_{\rm X}$ for $\alpha^{2}$\,CVn, is virtually ruled out.
It also appears unlikely that small differences in the magnetic topology of both stars lead to completely different wind channelling and wind shock configurations.
A more reasonable explanations would be to propose a different mass loss rate and that
IQ~Aur differs by intrinsic properties that promote mass loss and thus 
X-ray generation, for example it is more luminous and hotter than $\alpha^{2}$\,CVn.
In this case the X-ray generation would depend very sensitively on the underlying stellar parameter(s).
Alternatively, the MCWS phenomenon could be a transient one and currently only IQ~Aur is in an active wind phase with a high mass loss rate. 
Unfortunately, mass loss rates of Ap stars usually cannot be independently measured, but are determined via modeling of e.g. their X-ray properties.
However, the strong wind phase is believed to be connected to abundance anomalies, 
that are also observed in $\alpha^{2}$\,CVn.
It was already noticed by BM97, that any abundance anomalies should be removed by the rather high mass loss rate, 
thus requiring that IQ~Aur only recently, i.e. before a few Myr, entered the active wind phase. 
Although the details of the onset of the strong wind phase are rather unknown,
they estimate that Ap stars would spends a few percent of their lifetime in the strong wind phase, 
thus explaining that most Ap stars are no X-ray sources, at least at the moment.
Nevertheless, additional criteria might be necessarily fulfilled to make an A0p star a bright X-ray source.

\begin{figure}[t]
\includegraphics[width=90mm]{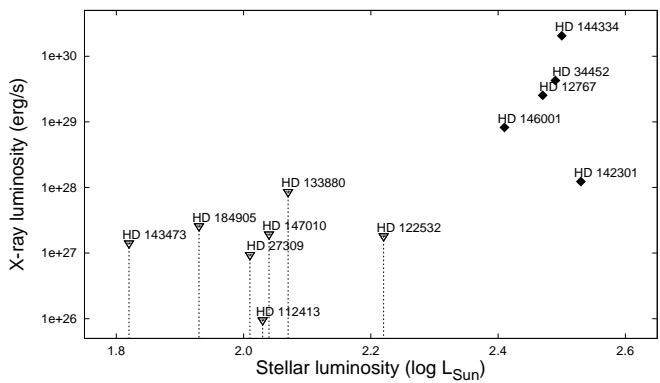}
\caption{\label{apcomp}X-ray vs. bolometric luminosity of the A0p stars IQ~Aur (HD 34452) and $\alpha^{2}$\,CVn (HD 112413) in comparison to other Ap/Bp stars. 
X-ray detected stars are plotted as diamonds, upper limits are indicated by open triangles.}
\end{figure}

\begin{table}[t!]
\setlength\tabcolsep{2.5pt}
\begin{center}
\caption{\label{tabapbp}X-ray observations of the comparison Ap/Bp stars.}
\begin{tabular}{llrrrr}
\hline\hline\\[-3mm]
Target & Sp.type & Dist. &Instr./Obs.ID & Dur.& $\log L_{\rm X}$\\
 & & (pc)  & &(ks) &(erg\,s$^{-1}$)\\\hline\\[-3mm]
HD 12767 &B9.5 Si &110 & ACIS-S (5391)& 3 & 29.4\\
HD 27309  & A0 SiCr &96 &XMM\,{\tiny (0201360201)}& 41 & $<$27.0\\
HD 122532 & B9 Si &169 & XMM\,{\tiny (0302900101)} & 130 &$<$27.3 \\
HD 133880 & B9 Si &126 &ACIS-I (2543) & 3 & $<$28.0\\
HD 142301 & B8IIIp& 139 & XMM\,{\tiny (0142630301)} & 20 & 28.1\\
HD 143473  & B9 Si &123 &ACIS-S (5385) & 15 &$<$27.2\\
HD 144334 &  B8 Si He-w& 149 &ACIS-S (5393) & 3 & 30.3\\
HD 146001 &B8 He-w &140&ACIS-S  (5394) & 3 & 28.9\\
HD 147010  &B9 SiCrFeSr&143 &ACIS-S  (5386) & 15 & $<$27.3\\
HD 184905 & A0 SiSrCr &165&ACIS-S (5387) & 15 &$<$27.4\\\hline
\end{tabular}
\end{center}
\end{table}

\subsection{X-rays from similar Ap/Bp stars}

\begin{figure}[t]
\hspace*{-0.3cm}\includegraphics[width=90mm]{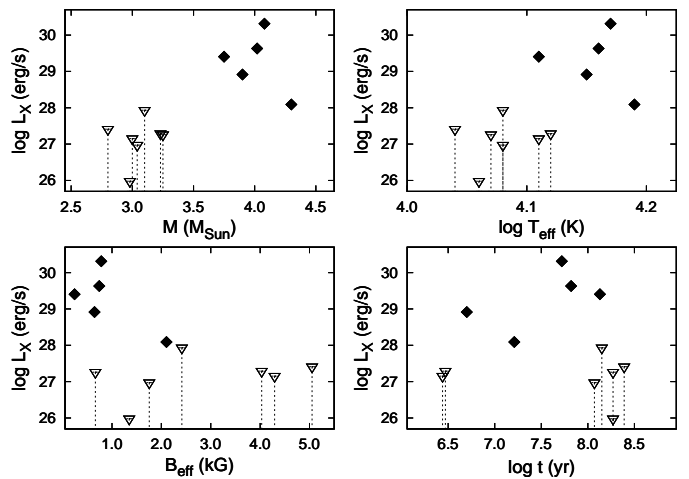}
\caption{\label{apcomp2}Correlations for early-Ap/late-Bp stars, $L_{\rm X}$ vs. stellar mass, eff. temperature, eff. magnetic field and age (symbols as in Fig.\,\ref{apcomp}).}
\end{figure}

Given the very different findings for IQ~Aur and $\alpha^{2}$\,CVn, we searched the {\it Chandra} and {\it XMM-Newton}
archives for observations of similar Ap/Bp stars, to derive some information on their general X-ray properties.
As a comparison sample we choose late-B to early-A magnetic CP stars with well determined distance and stellar parameter, that are either
likely single stars or spatially resolved in X-rays, although undiscovered binaries in the sample are not ruled out.
However, their contribution is likely minor, since even the overall fraction of close, i.e. spectroscopic, binaries in Ap stars of CP2/Si~type 
is with $\sim 20$\,\% significantly lower than those in non-magnetic stars \citep{abt73}.
The comparison sample is specified in Table~\ref{tabapbp};
some of these objects have already been presented, albeit with a different focus in \cite{ste06a,cze07}.
Most stars are Si-stars, but some of the hotter ones are also classified as He-weak, further
all stars have detected magnetic fields of a few hundred G up to several kG \citep{bych03}. The used
stellar data is again taken from \cite{koch06}, additional values are from \cite{lan07}. 
In Fig.\,\ref{apcomp} we show a comparison between the X-ray brightness and the stellar bolometric luminosity, 
errors are moderate, e.g. on $\log L_{\rm bol} \sim 0.1$.
The comparison sample reveals a striking similarity to our findings for IQ~Aur and $\alpha^{2}$\,CVn; either the stars are
X-ray bright with predominantly $\log L_{\rm X} \gtrsim 29.0$~erg\,s$^{-1}$ or X-ray faint with most upper limits being around $\log L_{\rm X} \lesssim 27.5$~erg\,s$^{-1}$, depending on the sensitivity of the respective exposure.
Only high luminosity objects are detected in X-rays, supporting an intrinsic X-ray generating mechanism. 
Furthermore, all stars with a high luminosity are detected as X-ray sources without exception.
Statistically the KS-probability of identical underlying distributions is well below 1\,\% for our sample.
A very similar dependence is present when using stellar mass, that is obviously related to luminosity; but
notably, we find that strength of the magnetic field or the stellar age do not play a major role in making an Ap/Bp star an X-ray source (see Fig.\,\ref{apcomp2}).
Unfortunately the sample does not cover the full parameter space, 
for example it lacks stars in the mass range 3.3\,--\,3.8~$M_{\odot}$ or with magnetic fields of $B_{\rm eff}>5$~kG and
its size is still quite small.
It thus does not allow to put strong constrains on the presence of a possible time-variability of the X-ray emission, but the fact that we detected
all stars above a stellar luminosity of around $200~L_{\odot}$ (or 3.6~$M_{\odot}$) and none below, 
suggests that a possible transient aspect of the MCWS phenomenon might be of lesser importance.
However, as is also obvious from Fig.\,\ref{apcomp} and Fig.\,\ref{apcomp2}, the X-ray luminosities of the detected stars have a large spread, indicating that other
parameters play an additional role in determining the specific X-ray properties of the individual stars.

\subsection{Where does the X-ray emission of IQ~Aur come from?}

The {\it XMM-Newton} observation clearly shows that the plasma temperatures of IQ~Aur are significantly higher than those estimated from the {\it ROSAT} data, 
where only weak indications for a hot component were found. This is very much relevant for the MCWS model,
since $T_{\rm X} \propto V_{\rm sh}^{2}$ and thus high wind speeds and a strong magnetic confinement must be present.
Temperatures of about 10~MK require typical shock velocities of several hundred up to 1000 km\,s$^{-1}$.
These values might be hard to obtain for A0p stars,
though relatively high shock temperatures would be a natural explanation for the absence or weakness of a plasma excess at 2~MK.
The relatively high \ion{O}{viii}/\ion{O}{vii} ratio in IQ~Aur could be explained by a quite high temperature in the colliding wind shocks of Ap stars; 
for example at shock speeds of 500\,--\,600~km\,s$^{-1}$ no \ion{O}{vii} excess
emission is produced at all. At a few stellar radii a line driven wind has already nearly its terminal speed and for strong magnetic confinement a large
longitudinal velocity component is expected to be present, easily explaining the absence of a cool excess even in a wind shock scenario.
Similarly, the X-ray luminosity of IQ~Aur requires a rather high mass loss rate and a very efficient conversion of kinetic wind energy to X-ray emission.
While not implausible, this requires IQ~Aur to drive a strong, fast wind.
However, given the presence of a disk, a fraction of the harder X-ray emission could be generated in minor reconnection events.
The \ion{O}{vii} $f/i$-ratio diagnostic indicates that the X-ray emitting plasma is not exposed to strong UV radiation.
Thus, if not shielded by an unknown medium, the X-ray emitting regions are not predominantly close to the stellar surface, 
but rather at distances of several stellar radii. 

Our lower limit derived of from the X-ray spectra of $d \approx 5$~$R_{*}$ matches the value derived above for the inner disk radius, therefore
a disk-like confinement region at distances of $5 - 15 $~$R_{*}$ is a suitable candidate for the X-ray emitting region.
On the other hand, the absence of significant rotational modulation over half of a rotation period
implies, that the changing viewing geometry introduces
a negligible amount of time-variable, intervening material in the line of sight, caused either by an optically thick disk or the star itself.
While the star as occulter appears unlikely given the large separation to the X-ray location, also for a disk rotational modulation is only expected
if it is rather thick and the viewing angle changes significantly with phase. A small inclination is unlikely, given the relatively large $V sini$ for an A0p star.
However, if the magnetic obliquity is small, rotational modulation effects are expected to be only minor.
In this case the inclination should be quite large and the system would be viewed nearly edge-on at all times.

A strong flare-like event with 100~MK plasma
is not expected to be produced by wind shocks even for magnetically channelled winds, thus these X-rays must be generated by magnetic reconnection.
X-ray emitting structures on the surface of IQ~Aur appear rather unlikely given the result from the X-ray diagnostics; additionally
the magnetic field of Ap/Bp stars is overall a large scale one and structurally differs from coronae where most 
X-ray plasma is confined in small-scale structures. Therefore in the MCWS scenario the flare requires the presence
of a rigidly rotating equatorial disk as predicted by the above outlined MHD simulations \citep{dou08}.
In this scenario magnetic reconnection occurs, leading to the production of hot X-ray emitting plasma and further, episodic
centrifugal breakouts of confined plasma produce flare-like events, 
originally proposed to explain X-ray flares on the Bp star $\sigma$~Ori (see discussion below) and \citep{dou06}.
In contrast to coronal flares, these magnetic reconnection events are driven by centrifugal mass ejections that disrupt outer disk regions.
During breakout, the centrifugal forces on the accumulated disk matter stretch the field lines more and more outwards 
and finally reconnection occurs. The event is accompanied by the ejection of outer disk mass, an inward propagating rebound and
releases sufficient energy to produce hard X-ray flares. The finding of metallicity changes during the flare implies 
that the accumulated material in the circumstellar disk, which is heated during the disruption,
is more metal rich compared to the plasma that generates the quasi-quiescent X-ray emission.
Our rough estimations of the size of the flaring structure would imply that the
outbreak-flare from IQ~Aur was a more local event in the disk, involving a large clump or segment rather than the entire disk.
Subsequent energy release or a change in the inner disk structure might be responsible for the flattening of the flare decay and
the continuously enhanced level of X-ray emission after the event. 
Overall this scenario appears suitable to also explain the flare event on IQ~Aur.
In this case, we have likely observed the first X-ray flare from an A0p star.

\subsection{X-ray bright companions for Ap/Bp stars?}

As an alternative explanation it has been postulated that X-ray detected intermediate mass stars 
possess unknown late-type companions that actually produce the X-ray emission.
Considering the estimated age of IQ~Aur, i.e. several ten up to hundred Myr
and comparing its X-ray luminosity with studies of the 70~Myr old Pleiades \citep{mic96}, an active
G or early-K dwarf as well as an M dwarf binary would be suitable candidate X-ray sources. 
However, young active stars have typically significantly higher plasma temperatures than observed for IQ~Aur \citep[see e.g.][]{gue04} and
their X-ray light curves show quite frequent X-ray flares of factor up to a few, 
whereas IQ~Aur is dominated by softer emission (Fig.\,\ref{coolex}) and exhibits except for the large flare only minor variability.
Other properties like the high $f/i$-ratio in the \ion{O}{vii} triplet could be easily explained with a companion, typical densities are usually low and
already a distance of 0.1~AU to IQ~Aur would be in sufficient to completely erase the influence of the UV-field.
Also the occurrence of a large flare are no problem for an active late-type companion
and the associated abundance changes are reminiscent of the chromospheric evaporation scenario describing the flare evolution. 

Another criterion to consider is the fact that IQ~Aur is also a bright radio source at 6\,cm
with $\log L_{6} \approx 16$~erg\,s$^{-1}$\,Hz$^{-1}$, whereas $\alpha^2$\,CVn with $\log L_{6} < 14.9$~erg\,s$^{-1}$\,Hz$^{-1}$ 
is radio dark \citep[radio properties from][]{lin92}.
Since a rather tight correlation between X-ray and radio brightness measured at $\nu \gtrsim 5$~GHz is present for coronal sources,
these  measurements allow to test the companion scenario by checking the well established X-ray/Microwave(radio) correlation (also known as G{\"u}del-Benz relation), 
which predicts $L_{\rm X}/L_{\rm R}=10^{15.5 \pm 0.5}$\,Hz for magnetically active stars \citep{gue93}.
With a quasi-quiescent X-ray luminosity of $\log L_{\rm X}\approx29.6$~erg\,s$^{-1}$,
IQ~Aur is radio-overluminous by
roughly two orders of magnitudes and thus significantly violates this correlation. The deviation is well in excess of the typical observed scatter 
that is typically  of a few.
Note that similar X-ray and radio fluxes were repeatedly observed for IQ~Aur, making intrinsic variability quite unlikely to be the reason for the discrepancy.
Also the B8p He-w/Si star HD~144334 (V\,929~Sco) with $\log L_{6} \approx 16.0$~erg\,s$^{-1}$\,Hz$^{-1}$ 
and $\log L_{\rm X} \approx 30.3$~erg/s violates the G{\"u}del-Benz relation by one to two orders of magnitude, 
similar to the B8V He-w star HD~146001 with a radio detection at $\log L_{6} \approx 16.0$~erg\,s$^{-1}$\,Hz$^{-1}$ (plus one U.L. at $\log L_{6} \lesssim 15.9$~erg\,s$^{-1}$\,Hz$^{-1}$).
An even stronger violation is observed for the B8IIIp He-w/Si star HD~142301 (V\,927~Sco), with $\log L_{6} \approx 17.1$~erg\,s$^{-1}$\,Hz$^{-1}$ 
and $\log L_{\rm X} \approx 28.1$~erg\,s$^{-1}$, this star is radio overluminous by more than four orders of magnitude.
On the other hand, all non X-ray detected comparison stars studied by \cite{lin92}, i.e. HD~122532, HD~147010, 
remained also undetected at radio wavelengths at a limit of $\log L_{6} \lesssim 15.7$~erg\,s$^{-1}$~Hz$^{-1}$.
This points to a joint occurrence of strong X-ray and radio emission for the sample stars. 
The fact that IQ~Aur and all other X-ray detected Ap/Bp stars violate the G{\"u}del-Benz relation by orders of magnitudes
suggests, that their emission does not originate from a late-type dwarfs, but a contributing companion in individual objects is not ruled out.

\subsection{What about hotter magnetic stars?}

The MCWS model has also been applied to a variety of hotter stars and it is illustrative to compare them with IQ~Aur.
A well studied example is the magnetic O5 star $\theta ^1$~Ori\,C, where \cite{gag05}
find that the dynamic MCWS model describes the X-ray properties of the star very well. It has an average X-ray luminosity
of $L_{\rm X} =1 \times 10^{33}$~erg\,s$^{-1}$ and
utilizing among other diagnostics $f/i$-line ratios, they conclude that the X-ray emitting plasma is located
at distances of only $0.2 \dots 0.5\,R_{*}$ rather close to the surface of the star.
$\theta ^1$~Ori\,C exhibits rotational modulated X-ray emission, since it is a tilted magnetic rotator with $i=45^{\circ}$ and $\beta =42^{\circ}$,
leading to a strong viewing angle dependent visibility and obscuration of the X-ray emitting torus.
$\theta ^1$~Ori\,C shows rather hard X-ray emission from very hot plasma at temperatures of up to 30\,MK, but no strong flaring.
Noteworthy, there have also been reports of X-ray flares on early type magnetic stars, a prominent example is the B2p star $\sigma$~Ori\,E.
While \cite{gro04} attribute a flare seen with {\it ROSAT} to $\sigma$~Ori\,E itself,
\cite{san04b} report a long duration ($>$10~h)
flare with a count rate increase by a factor of ten in an {\it XMM-Newton} observation and concluded that it more likely originates from a companion.
They derived a quasi-quiescent X-ray luminosity of $L_{\rm X} =9 \times 10^{30}$~erg\,s$^{-1}$ 
and plasma temperatures up to 10\,--\,15~MK for $\sigma$~Ori\,E.
Also non chemically peculiar, magnetic early-B stars are X-ray sources, e.g. the B0.2 star $\tau$~Sco
with  $L_{\rm X} =2 \times 10^{31}$~erg\,$s^{-1}$ \citep{coh03}. 
The {\it Chandra} spectra show emission lines that are broader than for coronal sources, but surprisingly narrow for a stellar wind shock scenario; together
with the presence of hard emission and X-ray plasma locations at several stellar radii make also $\tau$~Sco a candidate for MCWS generated X-ray emission.
Another example is the pulsating, hot magnetic giant star $\beta$~Cep, where \cite{fav09} again find a good agreement between the predictions from the
dynamic MCWS model and X-ray observations. It does again exhibit no flares and
with temperatures well below 10\,MK the X-ray emission is rather soft, despite its X-ray brightness of $\log L_{\rm X} =4 \times 10^{30}$~erg\,s$^{-1}$.
They derive distances of $\approx 5 R_{*}$ as the most likely location of the cooler X-ray emitting material, while the hotter plasma is likely closer to the star.
Since $\beta$~Cep has $i=45$ and $\beta =85^{\circ}$ but shows only minor modulation in its X-ray light curve,
they exclude the presence of a thick disk and attribute the X-ray emission exclusively to magnetically channelled wind shocks.

\begin{figure}[t]
\includegraphics[height=90mm,angle=-90]{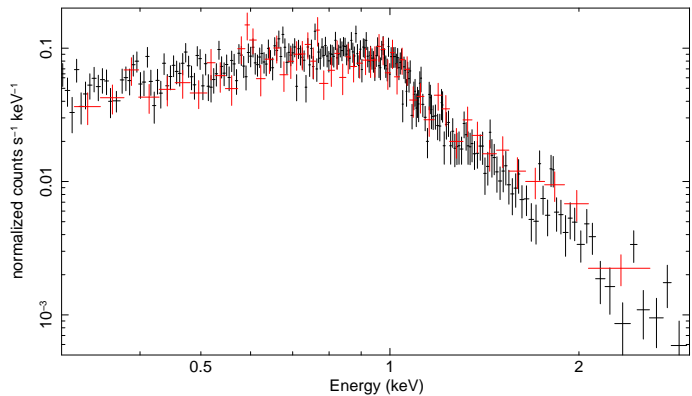}
\caption{\label{iqsig}Comparison of the quasi-quiescent X-ray spectra of the A0p star IQ~Aur (black) and the B2p star $\sigma$~Ori\,E (red) using XMM/PN data.}
\end{figure}

When comparing IQ Aur to the above mentioned massive magnetic stars, we find some similarities and overall trends.
Clearly, in general the X-ray luminosity decreases towards less massive stars, however all magnetic stars are with 
$\log L_{\rm X}/L_{\rm bol} \approx -6$ X-ray overluminous compared to similar 'normal' massive stars.
The bulk of the X-ray emitting plasma is located further away from the central star going from O over B to A-type stars 
and likely in the direction of increasing $\eta_ {*}$.
IQ~Aur is also by far the coolest object and the only star in this comparison, where the $f/i$ ratio of \ion{O}{vii} is high.
Comparing the X-ray temperatures, we find that the plasma on IQ~Aur is cooler than on $\theta ^1$~Ori\,C, comparable to $\sigma$~Ori\,E, but hotter than on $\beta$~Cep.
For stars with a moderate wind speeds well below 1000~km\,s$^{-1}$, 
in the MCWS scenario the lower temperature plasma ($\lesssim 5$~MK) is predominantly produced by wind shocks, 
whereas hotter plasma ($\gtrsim 10$~MK) is generated by reconnection in the disk. Notably, the strong flaring sources IQ~Aur and $\sigma$~Ori\,E have by far the
largest confinement parameter ($\log \eta_{*} \sim 6 \dots 7$), favoring the effective accumulation of matter in the equatorial region 
and consequently breakout events. As shown in Fig.\,\ref{iqsig}, also their quasi-quiescent spectra are very similar, despite the fact that
underlying stars differ significantly and $\sigma$~Ori\,E has a twenty times higher X-ray luminosity.

\subsection{The Ap/Bp star X-ray puzzle}

Our findings provide some clear hints for the solution to the X-ray puzzle of A0p stars. 
We do not find strong support for the hypothesis that the X-ray emission of Ap/Bp stars is 
produced by X-ray emitting companions, but cannot rule out that in some cases multiple sources contribute to the observed emission.
It is much more likely that the X-rays originate from intrinsic emission mainly generated in magnetically channelled wind shocks. However
predominantly or even exclusively the more massive, high luminosity stars appear as X-ray sources.
These late B/early A stars drive a stellar wind with a sufficiently large mass loss rate, whereas
further stellar properties like the magnetic topology or orientation and transient phenomena influence primarily the details of their X-ray appearance.
In the case of very strong magnetic confinement an equatorial disk structure forms, 
where magnetic reconnection and centrifugal breakout events occur that produce hard X-ray emission and occasional flare-like events.

Dedicated modeling of A0p stars and their magnetically channelled winds, companion searches, as well as more 
extensive studies of similar objects are highly desired to gain even deeper insights in the origin of
X-ray emission in these remarkable stars.

\section{Summary and conclusions}
\label{sum}

   \begin{enumerate}
\item We detect X-ray emission from IQ~Aur at $\log L_{\rm X}=29.4$~erg\,s$^{-1}$, but failed to detect 
$\alpha^2$\,CVn at $\log L_{\rm X}< 26$~erg\,s$^{-1}$ although both are A0p stars.
The quasi-quiescent spectrum of IQ~Aur exhibits significant emission from intermediate temperature and even 
hot plasma ($\gtrsim 10$~MK), in addition to a strong cool component at 3~MK. 
The UV sensitive \ion{O}{vii} $f/i$-ratio is high, thus the X-ray emission very likely originates from regions at
distances of $\gtrsim 5 R_{*}$ above the stellar surface. The X-ray light curve shows minor variability without clear signs of a viewing angle dependent rotational modulation. 

\item A massive X-ray flare with rise time of about 1~ks and fast decay occurred during the IQ~Aur observation.
The X-ray luminosity at the flare peak is about $\log L_{\rm X}=31.5$~erg\,s$^{-1}$ and plasma temperatures approach 100~MK.
The flare originates in a rather compact region and increases the metallicity of the X-ray emitting plasma significantly. 

\item Our X-ray data of IQ Aur can be reconciled within a dynamic MCWS model with strong magnetic confinement 
and build up of a rigid disk where magnetic reconnection events occur.
In this scenario, the flare-like events was produced by the centrifugal breakout of confined material. 
Alternatively, IQ~Aur might be a multiple system where likely both components contribute to the observed X-ray emission.

\item Magnetically channelled stellar winds are the most promising explanation for the X-ray detection of
Ap/Bp stars, although the presence of X-ray emission strongly depends on stellar parameters. 
A suitable quantity for setting the stage of MCWS generated X-rays appears to be a high stellar bolometric luminosity that supports a high mass loss, 
but magnetic topology, orientation or transient phenomena also play a role in determining the specific X-ray properties of an individual Ap/Bp star.
   
\end{enumerate}

\begin{acknowledgements}
This work is based on observations obtained with Chandra and XMM-Newton.
J.R. acknowledges support from DLR under 50QR0803.

\end{acknowledgements}

\bibliographystyle{aa}
\bibliography{16843}

\begin{thebibliography}{41}
\expandafter\ifx\csname natexlab\endcsname\relax\def\natexlab#1{#1}\fi

\bibitem[{{Abt} \& {Snowden}(1973)}]{abt73}
{Abt}, H.~A. \& {Snowden}, M.~S. 1973, \apjs, 25, 137

\bibitem[{{Arnaud}(1996)}]{xspec}
{Arnaud}, K.~A. 1996, in ASP Conf. Ser. 101: Astronomical Data Analysis
  Software and Systems V, ed. G.~H. {Jacoby} \& J.~{Barnes}, 17

\bibitem[{{Babel} \& {Montmerle}(1997)}]{bab97}
{Babel}, J. \& {Montmerle}, T. 1997, \aap, 323, 121

\bibitem[{{Berghoefer} {et~al.}(1997){Berghoefer}, {Schmitt}, {Danner}, \&
  {Cassinelli}}]{berg97}
{Berghoefer}, T.~W., {Schmitt}, J.~H.~M.~M., {Danner}, R., \& {Cassinelli},
  J.~P. 1997, \aap, 322, 167

\bibitem[{{Bohlender} {et~al.}(1993){Bohlender}, {Landstreet}, \&
  {Thompson}}]{boh93}
{Bohlender}, D.~A., {Landstreet}, J.~D., \& {Thompson}, I.~B. 1993, \aap, 269,
  355

\bibitem[{{Bychkov} {et~al.}(2003){Bychkov}, {Bychkova}, \& {Madej}}]{bych03}
{Bychkov}, V.~D., {Bychkova}, L.~V., \& {Madej}, J. 2003, \aap, 407, 631

\bibitem[{{Cohen} {et~al.}(2003){Cohen}, {de Messi{\`e}res}, {MacFarlane},
  {Miller}, {Cassinelli}, {Owocki}, \& {Liedahl}}]{coh03}
{Cohen}, D.~H., {de Messi{\`e}res}, G.~E., {MacFarlane}, J.~J., {et~al.} 2003,
  \apj, 586, 495

\bibitem[{{Czesla} \& {Schmitt}(2007)}]{cze07}
{Czesla}, S. \& {Schmitt}, J.~H.~H.~M. 2007, \aap, 465, 493

\bibitem[{{Favata} {et~al.}(2009){Favata}, {Neiner}, {Testa}, {Hussain}, \&
  {Sanz-Forcada}}]{fav09}
{Favata}, F., {Neiner}, C., {Testa}, P., {Hussain}, G., \& {Sanz-Forcada}, J.
  2009, \aap, 495, 217

\bibitem[{{Gagn{\'e}} {et~al.}(2005){Gagn{\'e}}, {Oksala}, {Cohen}, {Tonnesen},
  {ud-Doula}, {Owocki}, {Townsend}, \& {MacFarlane}}]{gag05}
{Gagn{\'e}}, M., {Oksala}, M.~E., {Cohen}, D.~H., {et~al.} 2005, \apj, 628, 986

\bibitem[{{Grevesse} \& {Sauval}(1998)}]{grsa}
{Grevesse}, N. \& {Sauval}, A.~J. 1998, Space Science Reviews, 85, 161

\bibitem[{{Groote} \& {Schmitt}(2004)}]{gro04}
{Groote}, D. \& {Schmitt}, J.~H.~M.~M. 2004, \aap, 418, 235

\bibitem[{{G{\"u}del}(2004)}]{gue04}
{G{\"u}del}, M. 2004, \aapr, 12, 71

\bibitem[{{G{\"u}del} \& {Benz}(1993)}]{gue93}
{G{\"u}del}, M. \& {Benz}, A.~O. 1993, \apjl, 405, L63

\bibitem[{{G{\"u}nther} \& {Schmitt}(2009)}]{guenther09}
{G{\"u}nther}, H.~M. \& {Schmitt}, J.~H.~M.~M. 2009, \aap, 494, 1041

\bibitem[{{Hubrig} {et~al.}(2007){Hubrig}, {North}, \& {Sch{\"o}ller}}]{hub07a}
{Hubrig}, S., {North}, P., \& {Sch{\"o}ller}, M. 2007, Astronomische
  Nachrichten, 328, 475

\bibitem[{{Kochukhov} \& {Bagnulo}(2006)}]{koch06}
{Kochukhov}, O. \& {Bagnulo}, S. 2006, \aap, 450, 763

\bibitem[{{Kochukhov} \& {Wade}(2010)}]{koch10}
{Kochukhov}, O. \& {Wade}, G.~A. 2010, \aap, 513, A13+

\bibitem[{{Landstreet}(1992)}]{lan92}
{Landstreet}, J.~D. 1992, \aapr, 4, 35

\bibitem[{{Landstreet} {et~al.}(2007){Landstreet}, {Bagnulo}, {Andretta},
  {Fossati}, {Mason}, {Silaj}, \& {Wade}}]{lan07}
{Landstreet}, J.~D., {Bagnulo}, S., {Andretta}, V., {et~al.} 2007, \aap, 470,
  685

\bibitem[{{Linsky} {et~al.}(1992){Linsky}, {Drake}, \& {Bastian}}]{lin92}
{Linsky}, J.~L., {Drake}, S.~A., \& {Bastian}, T.~S. 1992, \apj, 393, 341

\bibitem[{{Mewe} \& {Schrijver}(1978)}]{mewe78}
{Mewe}, R. \& {Schrijver}, J. 1978, \aap, 65, 99

\bibitem[{{Micela} {et~al.}(1996){Micela}, {Sciortino}, {Kashyap}, {Harnden},
  \& {Rosner}}]{mic96}
{Micela}, G., {Sciortino}, S., {Kashyap}, V., {Harnden}, Jr., F.~R., \&
  {Rosner}, R. 1996, \apjs, 102, 75

\bibitem[{{Ness} \& {Wichmann}(2002)}]{cora}
{Ness}, J.-U. \& {Wichmann}, R. 2002, Astron. Nachr., 323, 129

\bibitem[{{Porquet} {et~al.}(2001){Porquet}, {Mewe}, {Dubau}, {Raassen}, \&
  {Kaastra}}]{por01}
{Porquet}, D., {Mewe}, R., {Dubau}, J., {Raassen}, A.~J.~J., \& {Kaastra},
  J.~S. 2001, \aap, 376, 1113

\bibitem[{{Preston}(1974)}]{pres74}
{Preston}, G.~W. 1974, \araa, 12, 257

\bibitem[{{Reale}(2007)}]{rea07}
{Reale}, F. 2007, \aap, 471, 271

\bibitem[{{Reale} {et~al.}(1997){Reale}, {Betta}, {Peres}, {Serio}, \&
  {McTiernan}}]{rea97}
{Reale}, F., {Betta}, R., {Peres}, G., {Serio}, S., \& {McTiernan}, J. 1997,
  \aap, 325, 782

\bibitem[{{Robrade} \& {Schmitt}(2007)}]{rob07b}
{Robrade}, J. \& {Schmitt}, J.~H.~M.~M. 2007, \aap, 473, 229

\bibitem[{{Robrade} \& {Schmitt}(2009)}]{rob09a}
{Robrade}, J. \& {Schmitt}, J.~H.~M.~M. 2009, \aap, 497, 511

\bibitem[{{Sanz-Forcada} {et~al.}(2004){Sanz-Forcada}, {Franciosini}, \&
  {Pallavicini}}]{san04b}
{Sanz-Forcada}, J., {Franciosini}, E., \& {Pallavicini}, R. 2004, \aap, 421,
  715

\bibitem[{{Schmitt}(1997)}]{schmitt97}
{Schmitt}, J.~H.~M.~M. 1997, \aap, 318, 215

\bibitem[{{Schmitt} {et~al.}(1985){Schmitt}, {Golub}, {Harnden}, {Maxson},
  {Rosner}, \& {Vaiana}}]{schmitt85}
{Schmitt}, J.~H.~M.~M., {Golub}, L., {Harnden}, Jr., F.~R., {et~al.} 1985,
  \apj, 290, 307

\bibitem[{{Schmitt} \& {K{\"u}rster}(1993)}]{schmitt93}
{Schmitt}, J.~H.~M.~M. \& {K{\"u}rster}, M. 1993, Science, 262, 215

\bibitem[{{Schr{\"o}der} \& {Schmitt}(2007)}]{schroeder07}
{Schr{\"o}der}, C. \& {Schmitt}, J.~H.~M.~M. 2007, \aap, 475, 677

\bibitem[{{Stelzer} {et~al.}(2006){Stelzer}, {Hu{\'e}lamo}, {Micela}, \&
  {Hubrig}}]{ste06a}
{Stelzer}, B., {Hu{\'e}lamo}, N., {Micela}, G., \& {Hubrig}, S. 2006, \aap,
  452, 1001

\bibitem[{{Stelzer} {et~al.}(2009){Stelzer}, {Robrade}, {Schmitt}, \&
  {Bouvier}}]{ste09}
{Stelzer}, B., {Robrade}, J., {Schmitt}, J.~H.~M.~M., \& {Bouvier}, J. 2009,
  \aap, 493, 1109

\bibitem[{{Townsend} \& {Owocki}(2005)}]{tow05}
{Townsend}, R.~H.~D. \& {Owocki}, S.~P. 2005, \mnras, 357, 251

\bibitem[{{ud-Doula} \& {Owocki}(2002)}]{dou02}
{ud-Doula}, A. \& {Owocki}, S.~P. 2002, \apj, 576, 413

\bibitem[{{ud-Doula} {et~al.}(2008){ud-Doula}, {Owocki}, \& {Townsend}}]{dou08}
{ud-Doula}, A., {Owocki}, S.~P., \& {Townsend}, R.~H.~D. 2008, \mnras, 385, 97

\bibitem[{{ud-Doula} {et~al.}(2006){ud-Doula}, {Townsend}, \& {Owocki}}]{dou06}
{ud-Doula}, A., {Townsend}, R.~H.~D., \& {Owocki}, S.~P. 2006, \apjl, 640, L191

\end{thebibliography}

\end{document}